\documentclass[11pt,twoside]{article}

\usepackage{asp2006}
\usepackage{epsfig}
\usepackage{graphicx}

\begin{document}
\title{Halo Mergers, Galaxy Mergers, and Why Hubble Type Depends on Mass}
\author{Ariyeh H. Maller}
\affil{Physics Dept., New York City College of Technology, New York, NY}

\begin{abstract}
In the CDM cosmological framework structures grow from merging
with smaller structures.  Merging should have observable effects 
on galaxies including destroying disks and creating spheroids.  This 
proceeding aims to give a brief overview of how mergers occur in 
cosmological simulations. In this regard it is important to understand that dark matter halo mergers are not galaxy mergers; a theory of galaxy formation is necessary to connect the two.  Mergers of galaxies in hydrodynamical simulations show a stronger dependence on mass than halo mergers in N-body simulations.    If one knows how to connect galaxies to dark matter halos then the halo merger rate can be converted into a galaxy merger rate.  When this is done it becomes clear that major mergers are many times more common in more massive galaxies offering a possible explanation of why Hubble type depends on galaxy mass.
\end{abstract}

\section{Introduction}
Structure growth via mergers is one of the main predictions of CDM type cosmologies.  However, what is predicted is the merger rates of dark matter
halos, which are not directly observable.  Using dark matter halo merger rates to predict galaxy merger rates requires a theory of galaxy formation or at least a model of how galaxies populate dark matter halos.  In a similar way, what can actually be observed are close galaxy pairs, disturbed galaxies, or morphological differences between galaxies, all of which can only be indirectly tied to galaxy mergers using theoretical models. Thus connecting theory to observations poses a number of difficulties which are often not given enough attention.  In particular the halo merger rate is often used as an indicator of galaxy merger rates.   If galaxy mass scaled linearly with dark matter halo mass then this could possibly be true. But differences in the shapes of the galaxy stellar mass and halo mass functions imply that galaxy formation is much less efficient in low and high mass halos.  Thus we should expect that galaxy merger statistics should differ from halo merging statistics.  

\section{Dark Halo Mergers}
The majority of theoretical studies of merger rates analyze mergers of 
dark matter halos in N-body simulations \citep[][and references therein]{stew:07,fm:07,nd:07,pch:07}.  While there has been no study comparing the results of different analysis, differing treatments at least show qualitative agreement.  A summary of the results from these studies for halos associated with galaxy are:\\ 
\noindent 1. Halos rarely have major (greater than 1:3) mergers.\\
\noindent 2. Minor mergers (of order 1:10) are very common.\\
\noindent 3. The merger rate shows weak dependance on halo mass.\\

\begin{figure}   
\plottwo{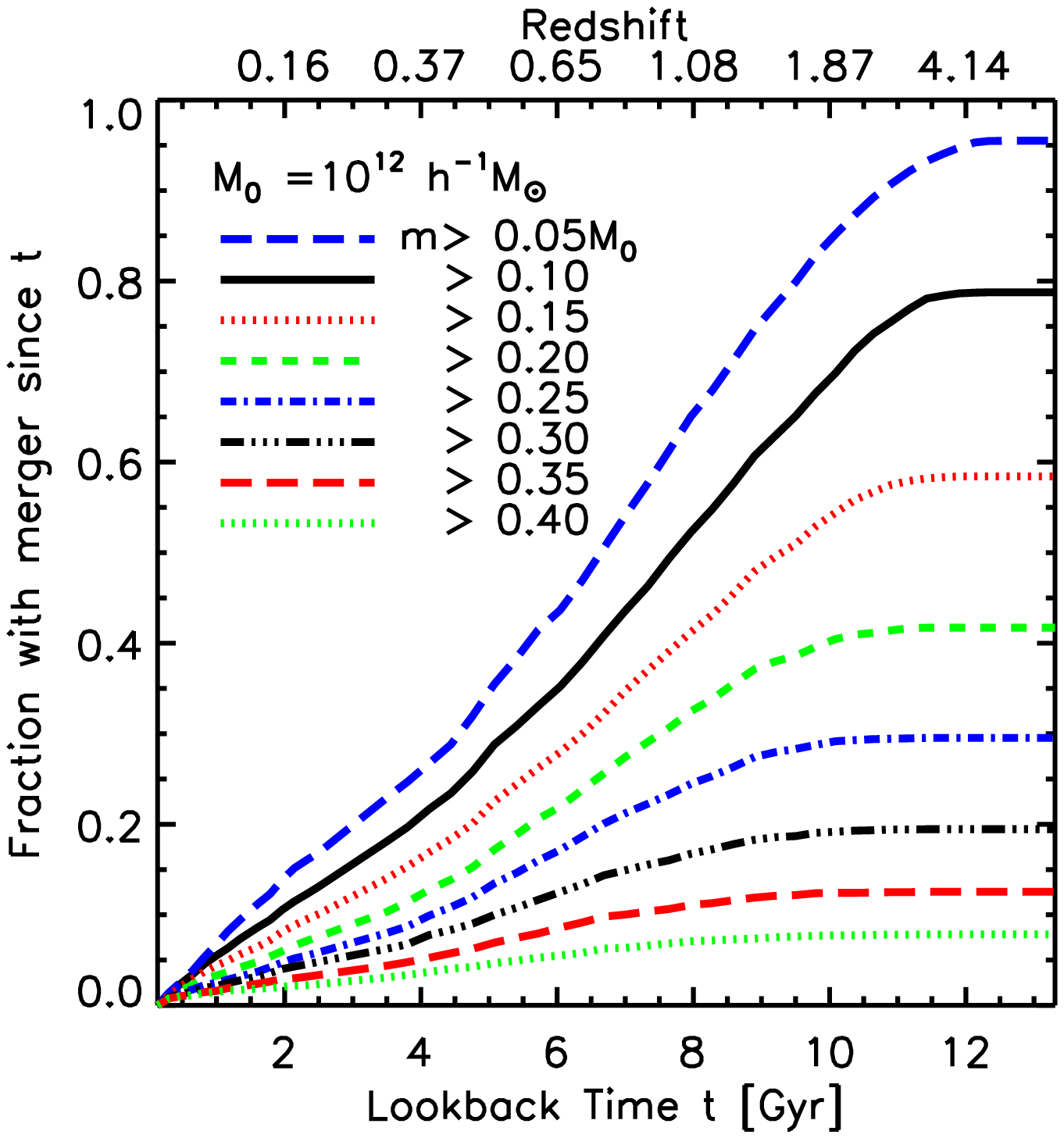}{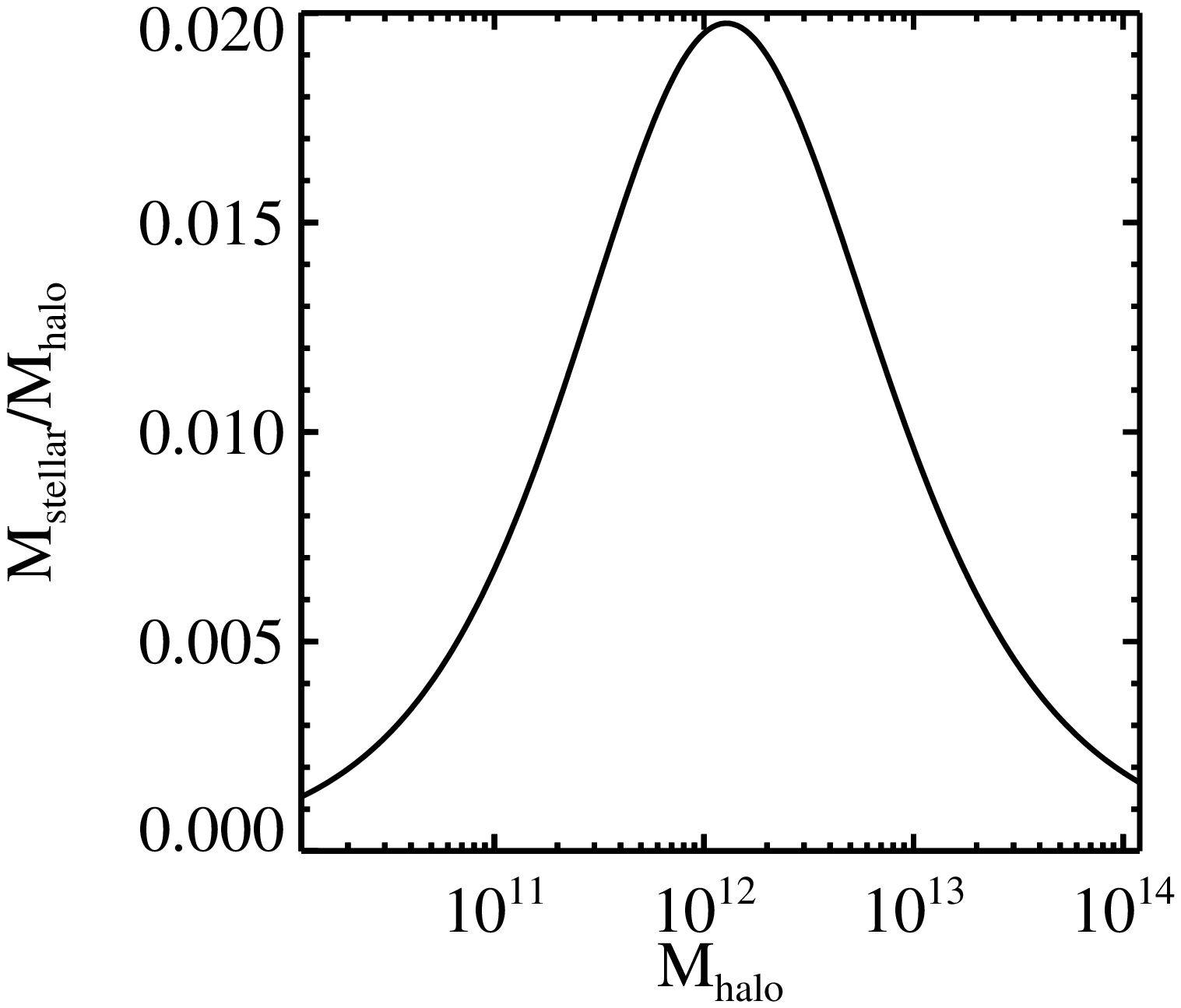}
\vspace{-0.4cm}
\caption{The left panel (Figure 5 from \citeauthor{stew:07}) shows the fraction of dark matter halos that have had a merger event greater than the 
given mass as a function of lookback time.  One sees that major mergers are rare, only about $30\%$ of  halos have accreted an object more massive than a fourth of their final mass.
On the other hand, minor mergers are common $80\%$ of halos have 
accreted something with more than one tenth of their final mass. 
The right panel shows the ratio of central galaxy stellar mass to halo mass as determined
by \citet{ymv:07} equation 7.  One sees that the efficiency of galaxy formation 
peaks a little above $10^{12} M_{\odot}$ and rapidly decreases away from that value.
}\label{fig:time}
\end{figure}

\noindent
These results are displayed in the left panel of Figure \ref{fig:time} taken 
from \citet{stew:07} which shows the fraction of halos that have accreted 
an object of a given mass as a function of lookback time.  Only about a third
of halos have had a major merger event involving a sizable amount of the halos final mass; however, $80\%$ of halos have had a merger with an object with a mass one tenth of the halo's final mass.  Creating this plot for different final halo masses results in almost no change aside from a very slight increase in the likelihood of a merger for all merger ratios.    

\section{Galaxy Mergers}
To go from dark matter halo merger rates to galaxy merger rates requires
a theory of galaxy formation.  Unfortunately at this time we have no 
theory that matches all the observed properties of galaxies, so the best 
that can be done is to explore the predictions of a given model of galaxy formation.  One possibility is to study the merger rates of galaxies in hydrodynamical simulations \citep{mall:06,mura:02}.  However, one must keep in mind, that hydrodynamical simulations at this time do not produce the observed galaxy stellar mass function.  Mergers in a hydrodynamical simulation are in most ways similar to the results of dark matter halos.  Major mergers are rare.  However, the merger rate does seem to show a much stronger dependance on galaxy mass then it does on halo mass \citep[see][Figure 9]{mall:06}.  There is some small
difference in the kinematics of galaxies compared to dark matter halos, most notably in their dynamical friction time scales, but this is unlikely to be 
the primary source of this mass dependance.  A much more important effect is that stellar mass does not scale linearly with halo mass. This means that the mass ratio of a galaxy merger may vary greatly from the mass ratio of the halos in which the galaxies reside.

\begin{figure}[h]
   \centering
   \plotone{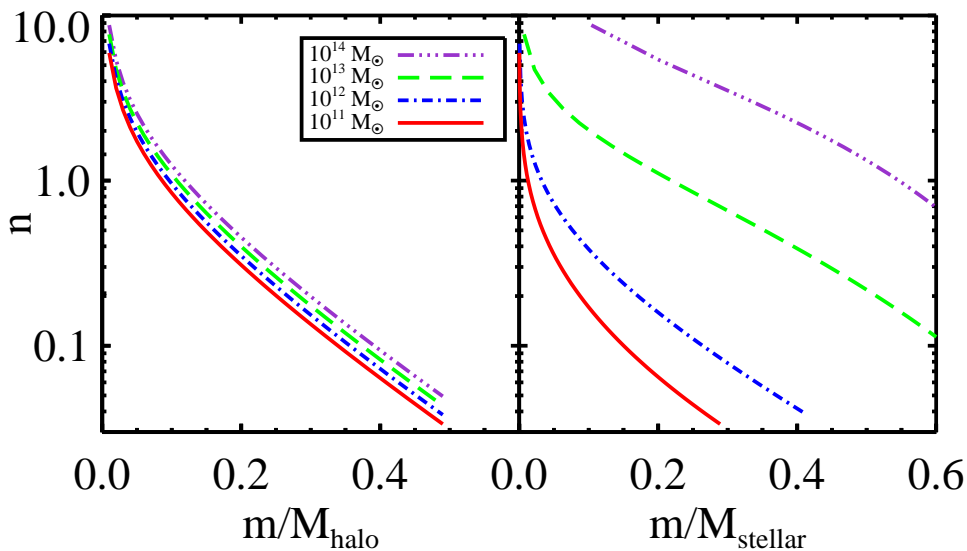} 
   \vspace{-0.7cm}
   \caption{The average number of objects with a mass of $m$ or greater
    that have been accreted over the age of the Universe in halos of different 
    masses .  The left panel shows 
    the functional fit of  \citet{stew:07} for four different final halo masses, 
    $M_{halo}$.  As emphasized in that paper, the dependance on mass is 
    rather weak and the four lines are close to identical.  The right panel shows
    the corresponding situation but now considering the stellar mass of the 
    central galaxy in the halo as determined by the conditional luminosity 
    function of \citet{ymv:07}.  Unlike in the case for dark matter halos 
    the distribution of masses that form a galaxy show a strong dependance 
    on the final galaxy mass.  In $10^{14} M_{\sun}$ halos the central galaxy
    is subjected to an average of one equal stellar mass galaxy merger, why in 
    $10^{11} M_{\sun}$ halos a galaxy usually doesn't even have one 
    merger with a twentieth of its mass. Note that in the right panel some of the
    lines extend past $0.5 m/M_{stellar}$ which means that the main progenitor 
    accretes a galaxy with a mass greater than half the final stellar mass.  This 
    is not possible if mass is conserved during mergers, which is probably not 
    the case as some stars are lost to inter-cluster light \citep{purc:07, chw:07}. 
    }\label{fig:comp}
\end{figure}

\section{Hubble Type}
This understanding can explain why Hubble type is such a strong function of galaxy mass.  A 1:3 merger in halo mass could result in a 1:10 or a 1:1 merger
in galaxy mass depending on how galaxies inhabit dark matter halos.  We don't 
know exactly how to assign galaxies to halos, but we know that galaxy formation must be very inefficient for high and low mass galaxies.  This can be seen in the 
right panel of Figure \ref{fig:time}, which shows the fraction of halo mass in 
the central galaxy using equation 7 of \citet{ymv:07}, which is obtained from a SDSS galaxy group catalogue.  While one can argue about the details of this
result, the generic shape of the function in the plot is well established.  Just from the shape of this function we can understand why Hubble type is a strong function of galaxy or halo mass.  For low mass halos the efficiency of galaxy formation increases with halo mass; so if two low mass halos merge the ratio of the stellar masses will be less than that of the halos.  But for high mass halos the efficiency of galaxy formation decreases with increasing mass, which leads to 
more nearly equal mass galaxy mergers.  This is illustrated in Figure \ref{fig:comp} which shows the mean number of objects accreted above a certain mass for different mass halos.  The left panel shows the dark matter case and
simply plots equation 3 from \citet{stew:07} for four different halo masses.  In 
comparison the right panel shows the same results for galaxy mass where the 
function from \citet{ymv:07} has been used to convert halo mass to central 
galaxy mass.  The point of this figure is just to show the striking difference in 
the two cases.  While there is almost no mass dependence in the dark matter
halo case, for galaxies the expected number of events can differ by almost two orders of magnitude.  Thus we would expect galaxy morphology to show
dependance on mass. 

In conclusion, Mergers of dark matter halos are largely independent of halo mass, but galaxy mergers are most likely very dependent on mass.  Measurements of galaxy merger statistics can be used as direct tests of galaxy formation models. 
While major mergers between halos are rather rare they can be relatively common between galaxies of certain masses depending on how galaxies inhabit dark halos.

\acknowledgements
AHM would like to acknowledge support to attend this conference from the PDAC at NYCCT. Also, the organizing committee is thanked for putting together an excellent and stimulating meeting.


\begin{thebibliography}

\bibitem[Conroy, Ho \& White (2007)]{chw:07}
Conroy, C., Ho, S., White, M. 2007,  \mnras, 379, 1491 

\bibitem[{{Fakhouri} \& {Ma}(2007)}]{fm:07}
{Fakhouri}, O. \& {Ma}, C.-P. 2007, ArXiv e-print/0710.4567

\bibitem[{{Maller} {et~al.}(2006){Maller}, {Katz}, {Kere{\v s}}, {Dav{\'e}}, \&
  {Weinberg}}]{mall:06}
{Maller}, A.~H., {Katz}, N., {Kere{\v s}}, D., {Dav{\'e}}, R., \& {Weinberg},
  D.~H. 2006, \apj, 647, 763

\bibitem[{Murali} {et~al.}(2002)]{mura:02}
{Murali}, C.,  {Katz}, N.,{Hernquist}, L., {Weinberg},D.~H., \& {Dav{\'e}}, R. 2002, \apj, 571, 1

\bibitem[{{Neistein} \& {Dekel}(2007)}]{nd:07}
{Neistein}, E. \& {Dekel}, A. 2007, ArXiv e-print/0708.1599

\bibitem[{{Parkinson} {et~al.}(2007){Parkinson}, {Cole}, \& {Helly}}]{pch:07}
{Parkinson}, H., {Cole}, S., \& {Helly}, J. 2007, ArXiv e-print/0708.1382

\bibitem[Purcell et~al. (2007)]{purc:07}
Purcell, C.~W., Bullock, J. S., \& Zentner, A. R. 2007, \apj, 666, 20

\bibitem[{{Stewart} {et~al.}(2007){Stewart}, {Bullock}, {Wechsler}, {Maller},
  \& {Zentner}}]{stew:07}
{Stewart}, K.~R., {Bullock}, J.~S., {Wechsler}, R.~H., {Maller}, A.~H., \&
  {Zentner}, A.~R. 2007, ApJ, submitted, ArXiv e-print/0711.5027

\bibitem[{{Yang} {et~al.}(2007){Yang}, {Mo}, \& {van den Bosch}}]{ymv:07}
{Yang}, X., {Mo}, H.~J., \& {van den Bosch}, F.~C. 2007, ArXiv e-print/0710.5096
\end{thebibliography}
\end{document}